\documentclass[pdflatex,sn-mathphys-num]{sn-jnl}% Math and Physical Sciences Numbered Reference Style

\usepackage{mathptmx}
\usepackage{graphicx}%
\usepackage{multirow}%
\usepackage{amsmath,amssymb,amsfonts}%
\usepackage{amsthm}%
\usepackage{mathrsfs}%
\usepackage[title]{appendix}%
\usepackage{xcolor}%
\usepackage{textcomp}%
\usepackage{manyfoot}%
\usepackage{booktabs}%
\usepackage{algorithm}%
\usepackage{algorithmicx}%
\usepackage{algpseudocode}%
\usepackage{listings}%
\usepackage[switch]{lineno}
\usepackage{caption}
\usepackage{bm}

\theoremstyle{thmstyleone}%
\theoremstyle{thmstyletwo}%

\theoremstyle{thmstylethree}%

\begin{document}
% \linenumbers

\title[
Emergence of unique hues from sparse coding of color in natural scenes]{\vspace{-5pt}Emergence of unique hues from sparse coding of color in natural scenes}

%%=============================================================%%
%% GivenName	-> \fnm{Joergen W.}
%% Particle	-> \spfx{van der} -> surname prefix
%% FamilyName	-> \sur{Ploeg}
%% Suffix	-> \sfx{IV}
%% \author*[1,2]{\fnm{Joergen W.} \spfx{van der} \sur{Ploeg} 
%%  \sfx{IV}}\email{iauthor@gmail.com}
%%=============================================================%%

\author*[1,2]{\fnm{Alexander} \sur{Belsten}}\email{belsten@berkeley.edu}

\author[3]{\fnm{E. Paxon} \sur{Frady}}\email{paxon.frady@gmail.com}

\author[1,2,4]{\fnm{Bruno A.} \sur{Olshausen}}\email{baolshausen@berkeley.edu}

%%=============================================================%%
%% Affiliations
%%=============================================================%%

\affil[1]{\orgdiv{Redwood Center for Theoretical Neuroscience}, 
\orgname{University of California, Berkeley}, 
\orgaddress{\city{Berkeley}, \state{CA}, \postcode{94720}, \country{USA}}}

\affil[2]{\orgdiv{Herbert Wertheim School of Optometry and Vision Science}, 
\orgname{University of California, Berkeley}, 
\orgaddress{\city{Berkeley}, \state{CA}, \postcode{94720}, \country{USA}}}

\affil[3]{\orgdiv{Neuromorphic Computing Lab}, 
\orgname{Intel}, 
\orgaddress{\city{Santa Clara}, \state{CA}, \postcode{95054}, \country{USA}}}

\affil[4]{\orgdiv{Helen Wills Neuroscience Institute}, 
\orgname{University of California, Berkeley}, 
\orgaddress{\city{Berkeley}, \state{CA}, \postcode{94720}, \country{USA}}}

\abstract{
    Our subjective experience of color is typically described by abstract properties such as hue, saturation, and brightness that do not directly correspond to sensory signals arising from cones in the retina. Along the hue dimension, certain colors---red, green, blue, and yellow---appear unique in that they are not perceived as a combination of other colors, and the pairs red-green and blue-yellow appear opposites. 
    However, the anatomical and physiological correlates of these `unique hues' within the brain and the reason for their existence remain a mystery. 
    Here, we demonstrate a direct connection between these hues and the statistics of the natural visual environment.
    Analysis of simulated cone responses on a dataset of 503 calibrated natural images reveals a strongly non-Gaussian distribution in 3D color space, with heavy tails in distinct, asymmetrically arranged directions. 
    A sparse coding model is then adapted to this data so as to minimize the total sum of coefficients on the basis vectors for representing the data.
    A six basis-vector model converges to the four unique hues in addition to black and white.
    Moreover, we find that the nonlinear nature of inference in the sparse coding model yields both excitatory and inhibitory interactions among latent variables; the former facilitates combining adjacent pairs of unique hues to encode intermediate hues situated between them, while the latter enforces mutual exclusivity between opposite unique hues. 
    Together, these findings shed new light on the distribution of color in the natural environment and provide a linking principle between this structure and the phenomenology of color appearance.
}

\maketitle

\section*{Introduction}\label{sec1}

In the late 19th century, 
% scientists sought to unravel the mechanisms underlying color perception in the eye and brain.
Ewald Hering and Hermann von Helmholtz engaged in a vigorous debate over their competing views regarding the mechanisms underlying color perception in the eye and brain.
While Helmholtz advanced a trichromatic theory of vision \cite{helmholtz1860optik}, Hering, relying largely on introspection, proposed a theory based on four unique hues---red, green, blue, and yellow---which in his view constituted atomistic elements of color appearance \cite{hering1878lehre}.
Noting that certain pairs of these hues are never simultaneously perceived in any color (i.e., red-green and blue-yellow are mutually exclusive), Hering showed how they could be organized into opponent pairs and serve as a perceptual basis for describing all other colors.
Ultimately, 
% much has been learned about the early stages of physiological color processing. 
the Helmholtz-Young trichromatic theory was linked to the presence of three types of cones in the retina, and subsequently, a cone-opponent encoding strategy was identified in the retina and lateral geniculate nucleus (LGN) whereby neurons compute differences in activation between cone types~\cite{krauskopf1982cardinal,derrington1984chromatic}.
At the same time, Hering's unique hues came to be adopted by psychology, vision science, and color technology as a canonical basis for describing color appearance~\cite{devaloisseeingcolorvision2000, fairchild2015color, wolfe2022perception}. 
However, the {\em cone-opponent} representation of the early visual system does not align with the {\em color-opponent} organization implied by the unique hues~\cite{de1997hue,webster2000variations,wuerger2005cone,wool2015salience}.  Attempts to find neurophysiological correlates of the unique hues
% This has led to an ongoing search for a neural basis for the unique hues in cortex; 
% however, clear physiological mechanisms demonstrating their privileged status have yet to be identified
have been largely inconclusive~\cite{stoughton2008neural,mollon2009neural,conway2023color}, with the exception of Forder \textit{et al.}~\cite{forder2017neural}.
Psychophysical evidence reliably distinguishing their `uniqueness' as compared to other hues also remains inconclusive~\cite{bosten2014empirical,conway2023color}.
Thus, the alleged privileged status of the unique hues remains both enigmatic~\cite{saunders1997there,mollon1997nature,valberg2001unique} and controversial~\cite{jameson199714,conway2023color}, as they do not neatly map onto any well understood property of light or the visual system.

It has long been suggested that color perception is shaped by the statistics of the natural visual environment~\cite{shepard1997perceptual,webster1997adaptation,mollon1997nature,mollon2006monge,rosenthal2018color,skelton2024effects,hedjar2025importance}. Could this provide an account for the unique hues?  
% Here, we approach this question by studying the statistics of color in the natural visual environment. 
Previous work by Webster and Mollon~\cite{webster1997adaptation} analyzed chromatic variation in natural images and found substantial variation along blue-yellow hue directions.
Ruderman \textit{et al.}~\cite{ruderman1998statistics} characterized pairwise correlations among simulated cone responses on calibrated hyperspectral images and found that a decorrelating basis computed through principal component analysis (PCA) corresponds to physiological cone-opponent mechanisms found in the retina and LGN~\cite{derrington1984chromatic}. 
Models that characterize higher-order statistics have been adapted to spatiochromatic natural images and shown to produce orientation- and chromatically-selective receptive fields~\cite{hoyer2000independent,wachtler2001chromatic,lee2002color,doi2003spatiochromatic, caywood2004independent, wachtler2007cone}. However, none of these studies resulted in bases aligned with the unique hues. One possible reason is that learning bases from spatiochromatic image patches mixes spatial and color statistics, obscuring higher-order structure that is specific to the color distribution itself.
Yendrikhovskij~\cite{yendrikhovskij2001computing} applied clustering algorithms to pixels of natural images in a perceptually uniform color space and found an alignment between the resulting clusters and color-naming lexicons, which points toward a relation to the unique hues.
However, one drawback of these studies is that they were based on particular, limited datasets (e.g., approximately $10^4$--$10^5$ pixels). More importantly, these approaches did not directly examine the full structure of the 3D color distribution.

\begin{figure*}[t]
    \centering
    \makebox[\textwidth][c]{\includegraphics[width=17.78cm]{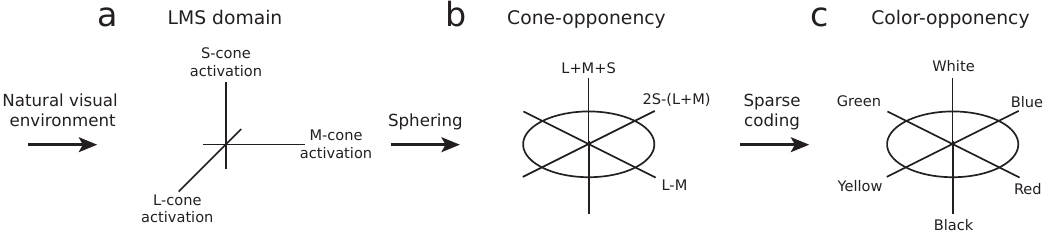}}
    \caption{
        Three-stage model of chromatic processing in the visual system.  
        \textbf{a})
            The pixels of a large dataset of natural images are represented in LMS cone activation space (stage~1).
        \textbf{b}) 
            A linear sphering transform decorrelates the activations by first rotating to the principal components,  yielding a set of {\em cone-opponent} bases, and then rescaling each axis to achieve unit variance in all directions (stage~2). 
        \textbf{c})
            A set of bases that efficiently represent higher-order structure in the cone-opponent space is derived using nonnegative sparse coding models (stage~3). 
            A six basis-vector model converges to directions that align with the unique hues, forming a new 6D, nonnegative space whose axes correspond to perceptual color categories. The inference process of sparse coding leads to mutual exclusivity between opponent colors (i.e., red-green, blue-yellow, black-white) depicted here in a 3D space with opposing colors at opposite ends of each axis.  Notably, these color-opponent axes, when projected back into the cone-opponent space, are non-orthogonal and not aligned with the cone-opponent axes. 
         Circles in \textbf{b} and \textbf{c} indicate chromatic planes. 
    }
    \label{fig:overview}
\end{figure*}

% paragraph 3
% Were going to look at the data. 
In this study, we analyze a large composite dataset comprising more than $10^8$ simulated cone activations sampled from images collected by different investigators and spanning a diverse range of visual environments.  
% composite dataset of simulated LMS cone activations in response to natural scenes assembled from five different, publicly available datasets. 
% We also studied each dataset individually; these results are presented in the Supplemental Information.
We then derive encoding models adapted to the structure of these data, following the three-stage pipeline depicted in Figure~\ref{fig:overview}.
Rather than examining the full spatiochromatic structure of image patches, we focus exclusively on the statistics of the three-dimensional color distribution defined by the joint activations of long-, medium-, and short-wavelength selective cones (henceforth referred to as LMS space).
Thus, our analysis treats pixels as samples from a color distribution and does not model the spatial organization of chromatic structure, such as through spatial correlations, edges, or object-level color structure. 
Extending the present approach to the joint statistics of color and spatial structure is an important but more complex problem that we defer for future study.

% paragraph 5
% Next, the data are transformed into a space that more directly captures this higher-order structure. Following previous work on natural image statistics that utilizes sparse coding to capture heavy-tailed structure, we adapt a sparse coding model to the data in the sphered color space.  
% Given the higher-order structure of the data, how should it be encoded and represented? 
% Due to the strong asymmetry in the data, we utilize nonnegative sparse coding models with varying degrees of overcompleteness. 
% These models use a set of basis vectors to efficiently represent non-Gaussian structure, and thus reflect the underlying statistical regularities of the color environment. 
% Our second finding is that a 6-basis-vector model learns a set of bases that align closely with the four unique hues plus black and white.
% The inference process in sparse coding leads to mutual exclusivity between opponent basis vectors, enabling the construction of a color-opponent representation (third stage, Fig.~\ref{fig:overview}c).
% Furthermore, increasing the model’s overcompleteness progressively refines the color basis in a systematic manner---mirroring how languages seem to subdivide color space into an expanding color lexicon~\cite{kay1999color, zaslavsky2018efficient}.

\section*{Results}\label{sec2}
The results below are based on a large composite dataset of simulated LMS cone activations in response to natural scenes assembled from five different, publicly available datasets. 
Details of each component dataset and the preprocessing procedures are provided in the Methods section. 
Briefly, the data from each image is rescaled independently for each cone type and then passed through a compressive nonlinearity.
% that effectively performs histogram equalization. 
Finally, the mean across all channels was subtracted so that the dataset was zero-centered before further processing.

The analysis proceeds through three stages. First, natural scenes are represented in LMS cone activation space (Fig.~\ref{fig:overview}a). 
Second, these responses are linearly decorrelated and sphered, yielding a cone-opponent coordinate system in which second-order structure has been removed and higher-order deviations from Gaussianity are made explicit (Fig.~\ref{fig:overview}b). 
Third, the sphered cone activations are represented using a nonnegative sparse coding model, which learns a basis adapted to the asymmetric, heavy-tailed structure of the distribution of color in natural scenes. Nonlinear inference in this model transforms each sphered cone activation into a sparse, nonnegative set of latent variables. In the six-basis-vector model, the learned basis vectors align with the unique hues, and the nonlinear inference dynamics produce mutual exclusivity between latent variables corresponding to opponent-color pairs (Fig.~\ref{fig:overview}c).

\subsection*{Principal components align with cone-opponent color space}

\begin{figure*}[!t]
    \makebox[\textwidth][c]{\includegraphics[width=16cm]{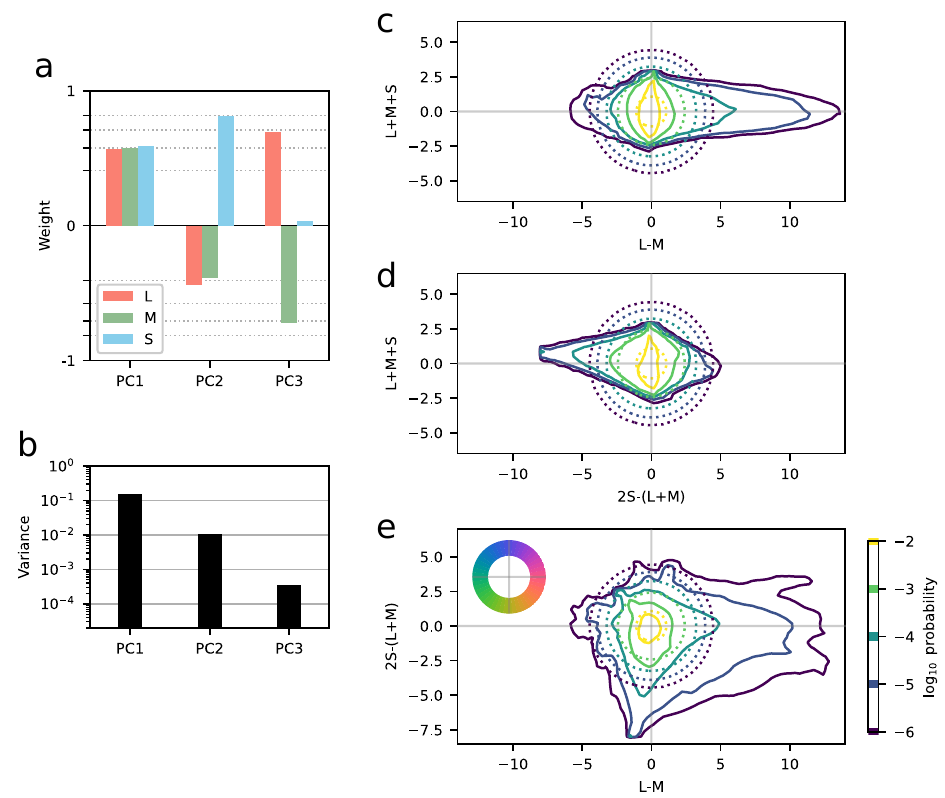}}
    \caption{ 
        \textbf{a}) 
            Principal components of the distribution of LMS activations. 
            The weights' alignment with the horizontal dashed lines at 
            $\pm\text{sqrt}(1/6)$, 
            $\pm\text{sqrt}(1/3)$, 
            $\pm\text{sqrt}(1/2)$,
            and $\pm\text{sqrt}(2/3)$
            reveals the near-integer-multiple relationship in how they mix L, M, and S signals (as pointed out by~\cite{ruderman1998statistics}).
        \textbf{b}) 
            Variance of each principal component.
        \textbf{c}-\textbf{e})
            LMS activation distributions in the sphered color space exhibit anisotropic, non-Gaussian structure. Solid lines show isoprobability contours after projection onto three different 2D planes defined by each possible pairing of axes in the sphered color space, as indicated by the axis labels. Dashed contours indicate isoprobability contours for a multivariate isotropic Gaussian distribution with the same variance as the data distribution. The contours are equally spaced in log-probability, using the same color scale for both solid and dashed contours. The color circle inset in \textbf{e} indicates the hue of each angle within the cone-opponent chromatic plane.
    }
    \label{fig:stats}
\end{figure*}

A natural first step in understanding the structure of multivariate data is to apply PCA, which transforms the data into a new space where variables are pairwise decorrelated~\cite{simoncelli2001natural} (see Methods).
Applying PCA to the composite dataset of simulated cone activations yields the principal components and corresponding variances shown in Figure~\ref{fig:stats}a,b (see also Fig.~S1).
The first principal component (PC1) sums the activations of L, M, and S cones, corresponding to an achromatic channel. The remaining two principal components compare the activations of different cone types: the second component (PC2) contrasts the activation of S cones with the net activation of L and M cones, and the third (PC3) compares the activation of L and M cones. 
These components align nearly identically with those obtained previously by Ruderman \textit{et al.}~\cite{ruderman1998statistics}. 
Notably, PC2 and PC3 correspond to the cone-opponent mechanisms which have been characterized psychophysically~\cite{krauskopf1982cardinal} and physiologically~\cite{derrington1984chromatic}.

% Notably, the principal components have a direct correspondence with the axes of the 3D Derrington, Krauskopf \& Lennie (DKL) cone-opponent color space \cite{derrington1984chromatic}. The first principal component aligns with the vertical axis of the DKL color space, while the second and third axes align with the cone-opponent axes that define the DKL chromatic plane. 

The PCA transformation also reveals a striking anisotropy in the data distribution: the variance on PC1 is more than 10-fold greater than on PC2, and there is an even greater difference in variance between PC2 and PC3 (Fig.~\ref{fig:stats}b). This strong anisotropy makes it difficult to ascertain the full 3D structure of the joint distribution. Therefore, we equalize the variance in all directions of the space by rescaling each principal component axis by the inverse square root of its corresponding variance (see Methods). Throughout the remaining sections of this paper, we will refer to the unit norm principal component vectors as PC1, PC2, and PC3, and their corresponding rescaled versions based on their predominant cone inputs: L+M+S, 2S-(L+M), and L-M, respectively. In addition, we refer to the 3D space constructed by the latter three axes as the sphered color space.

Our motivation for rescaling is twofold. First, for clarity of visualization: when examining the 3D distribution of activations in the decorrelated space, having axes with equal variance makes any deviation from Gaussianity readily apparent. Second, we conjecture that the brain may also rescale sensory data in this way. As argued by various authors~\cite{laughlin1981simple,field1987relations,graham2006can}, neural systems adjust their gain to fill their output dynamic range. Thus, neurons tuned to low-variance directions in the data exhibit comparatively higher gain than those sensitive to high-variance directions in the data. Since the achromatic component has substantially higher variance than the chromatic components (Fig.~\ref{fig:stats}b), amplification of chromatic signals relative to achromatic is necessary to achieve equal variance. 
This prediction aligns with psychophysical measurements of detection thresholds, which reveal substantially higher sensitivity to chromatic increments than to luminance increments~\cite{stromeyer1985second,wandell1985color}.

\subsection*{The distribution of LMS activations is highly non-Gaussian}

Figure~\ref{fig:stats}c-e shows the structure of the sphered distribution via three 2D projections of the full 3D distribution, using each possible pairing of L-M, 2S-(L+M), and L+M+S axes (see also Fig.~S2).
The resulting density of each projection is plotted in terms of its isoprobability contours.
For comparison, contours from an isotropic Gaussian distribution with identical variance are overlaid on each plot. 
Comparing the isoprobability contours of the LMS activations to those of the Gaussian reveals pronounced asymmetries and substantial deviations from Gaussianity.

First, consider the achromatic dimension in Figure~\ref{fig:stats}c,d. The low-probability contours of the LMS distribution are slightly more compact along the achromatic axis than the Gaussian contours, suggesting a mild sub-Gaussian shape. Quantitatively, this is confirmed by a negative excess kurtosis of -0.6 for 1D projection along the achromatic axis (compared to 0 for a Gaussian; see Table~S1).

By contrast, the chromatically sensitive dimensions (L-M and 2S-(L+M)) exhibit strong heavy-tailed structure as the isoprobability contours in these directions extend farther from the origin than those of a Gaussian. The excess kurtosis for 1D projections along these directions is 4.3 for 2S-(L+M), and 27.0 for L-M. Additionally, the tails are asymmetric, with greater probability mass in the extremes of the positive L-M direction than that of the negative. Similarly, there is more probability mass in the extremes of the negative 2S-(L+M) direction than in the positive. The pattern of heavy-tailed structure within the chromatic plane and slightly sub-Gaussian structure along the achromatic axis is consistent across the different LMS datasets analyzed (Table~S1).

Figure~\ref{fig:stats}e shows the joint distribution within the plane defined by the cone-opponent axes. The heavy-tailed structure varies substantially with azimuth angle (the angle formed between a point's projection onto the plane and the positive L-M direction). Certain directions, such as $0^\circ$ and $260^\circ$, display pronounced heavy tails, while others, such as $135^\circ$, exhibit more Gaussian or even sub-Gaussian tails. Notably, the structure is asymmetric between opposing directions (e.g., $0^\circ$ vs. $180^\circ$ and $135^\circ$ vs. $315^\circ$).

Early observations by Hendley and Hecht~\cite{hendley1949colors} reported that natural scenes generally lack saturated colors. While most pixel activations indeed cluster near the achromatic axis, the large dataset of over 225 million pixels reveals a richer structure: saturated colors actually occur with greater frequency than expected from a random process with the same variance as the data (i.e., a Gaussian distribution). Detecting this heavy-tailed structure requires large sample sizes; smaller datasets (e.g., tens of thousands of pixels) would fail to capture these tails accurately.

What gives rise to the heavy-tailed structure observed in the chromatic plane? It is not likely to be an artifact of the preprocessing pipeline, because rescaling and histogram equalization would, if anything, reduce kurtosis rather than amplify it~\cite{baddeley1996searching}. Moreover, the heavy-tailed directions do not show correspondence to the directions of the L, M and S cones in the sphered space, which point in $330^\circ$, $210^\circ$, and $90^\circ$ directions, respectively. The fact that the heavy-tailed structure survives after averaging over datasets collected by different investigators using different cameras suggests that it is a persistent property of the natural environment as opposed to idiosyncratic properties of any given dataset.  Previous work has shown that color in natural scenes tends to be spatially patchy, with large, relatively uniform regions compared to the more fine-grained variation of luminance~\cite{yoonessi2008color}. This spatial structure may arise because surface color is a material property that tends to remain coherent within an object, whereas luminance varies both with surface reflectance and with other factors such as shading and illumination geometry. Additionally, Rosenthal \textit{et al.}~\cite{rosenthal2018color} found that objects tend to have higher chromatic saturation, suggesting that the heavy tails in the chromatic distribution (Fig.~\ref{fig:stats}e) are driven by the presence of highly saturated, chromatically consistent objects (e.g., flora, foliage, sky, and fruit) within natural scenes. 
A hue-resolved analysis that links the heavy-tailed structure observed here to image content will be interesting to explore in future work.

\subsection*{Modeling higher-order structure}

While heavy-tailed structure is revealed by projection onto pairs of principal components, PCA per se cannot provide a model of this structure as it is computed solely from second-order statistics (the covariance matrix of the data), which are blind to such structure~\cite{field1994goal,simoncelli2001natural}. Thus, one must turn to a modeling approach capable of capturing higher-order statistics.

An alternative linear generative modeling approach capable of capturing such higher-order statistics is independent component analysis (ICA), or its highly related cousin, sparse coding~\cite{bell1995information,olshausen1996emergence,hyvarinen2009natural}. Intuitively, sparse coding seeks a set of basis vectors such that each data point can be reconstructed using only a small number of active coefficients. This sparsity-inducing penalty favors learning basis vectors which align with the directions in the data that occur rarely but with high magnitude, i.e., high-kurtosis or heavy-tailed directions~\cite{field1994goal,simoncelli2001natural,ganguli2012compressed}.
Sparse coding applied to natural images provides a canonical example of this principle.
When adapted to such data, the learned basis aligns itself with heavy-tailed directions in image space and exhibits receptive field properties similar to those of simple cells in primary visual cortex~\cite{olshausen1997sparse}. 
Theoretical and computational studies have also demonstrated the utility of sparse representations: sparsity makes complex structure in sensory inputs explicit, improves storage capacity, enables easier readout by high-order neural populations, and produces an energy-efficient representation~\cite{olshausen2004sparse}.

The sparse coding generative model assumes each 3D data vector $\mathbf{x}$ to be described as a linear combination of $m$ unit-norm basis vectors $\mathbf{a}_1,\mathbf{a}_2, ..., \mathbf{a}_m$ plus Gaussian noise $\bm{\varepsilon}$:
\vspace{-1pt}
\begin{align}
    \mathbf{x} = \sum_{i=1}^m\mathbf{a}_i{s}_i + \bm{\varepsilon}, \label{eqn:generative_model}
\end{align}
\vspace{-1pt}
The variance of the noise $\bm{\varepsilon}$ is assumed to be small relative to that captured by the basis vectors $\mathbf{a}_i$.  Importantly, there is no assumption of orthogonality on the $\mathbf{a}_i$, and they can even be overcomplete and hence linearly dependent, providing greater flexibility in capturing structure in the data.
Sparsity in the latent variables $s_i$ is enforced via a factorial prior that encourages their values to be zero. Typically, the prior is chosen to be symmetric about zero (e.g., a Laplacian distribution), enabling the $s_i$ to take on both positive and negative values. Given the asymmetric structure evident in the data, we impose a nonnegativity constraint on their values (i.e., $s_i\geq0 \ \text{for all}\  i$).

While the sparse coding model is a linear generative model, the latent variables $s_1,s_2,...,s_m$ are a nonlinear function of the data $\mathbf{x}$. The $s_i$ are computed by minimizing an energy function $E$ subject to a nonnegativity constraint:
\begin{align}
    \text{minimize} & \quad E = \frac{1}{2}\Big\lVert\mathbf{x} - \sum_{i=1}^m\mathbf{a}_i{s}_i\Big\rVert_2^2\; +\; \lambda \sum_{i=1}^m s_i \label{eqn:energy} \\
\text{subject to} & \quad s_i \geq 0 \quad\text{for} \ i = 1,2,...,m, \label{eqn:nonneg_constraint} 
\end{align}
where parameter $\lambda$ controls the trade-off between reconstruction quality and sparsity. 
This minimization corresponds to performing maximum-a-posteriori (MAP) Bayesian inference, $\mathbf{s}=\arg\max_{\mathbf{s}'} p(\mathbf{s}'|\mathbf{x},\mathbf{A})$, where vector $\mathbf{s}$ has elements $s_i$ and matrix $\mathbf{A}$ has columns $\mathbf{a}_i$~\cite{olshausen1997sparse}.

The optimization problem of equations~\ref{eqn:energy} and \ref{eqn:nonneg_constraint} is solved using the Locally Competitive Algorithm (LCA; see Methods). 
The basis vectors $\mathbf{a}_i$ are then adapted to the structure of the data by gradient descent on $E$ averaged over many data samples, using the MAP-inferred $\mathbf{s}$ for each data sample $\mathbf{x}$ (see Methods). Note that the nonnegativity assumption limits the minimum number of basis vectors to $m=4$, as at least four vectors are necessary to span three dimensions under nonnegative weightings.

\subsection*{Progression of basis vector solutions with increasing overcompleteness}

\begin{figure*}[t]
    \centering
    \makebox[\textwidth][c]{\includegraphics[width=17.78cm]{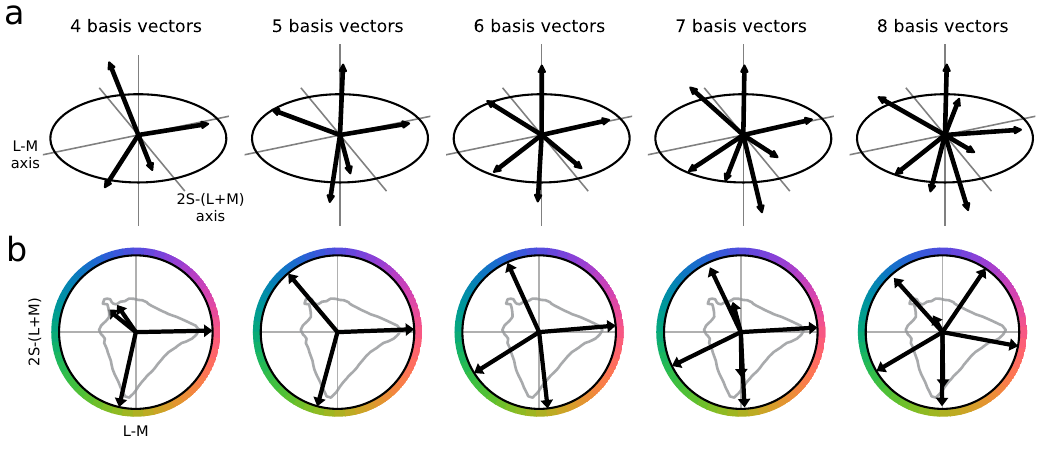}}
    % \vspace{-1cm}
    \caption{
        \textbf{a}) 
            Adapted nonnegative sparse coding basis vectors are shown in the sphered color space. The number of basis vectors is varied from left to right from four to eight. Each basis vector has unit length. The black ellipse depicts a unit-radius circle within the chromatic plane. 
        \textbf{b})
            Basis vectors from \textbf{a} are shown projected into the chromatic plane. The black circle depicts unit radius, and the outer ring shows the corresponding color for each hue angle. The gray curve reproduces the $\log_{10}\text{probability}=-4$ contour from Figure~\ref{fig:stats}e; it is uniformly rescaled for visual comparison, so shape is preserved but absolute scale is not. Vectors closely aligned with the achromatic axis project to (near) zero and may not be visible.
    }
    \label{fig:sc_models}
\end{figure*}

Figure~\ref{fig:sc_models} shows the basis vectors $\mathbf{a}_i$ that emerge from nonnegative sparse coding models trained on the data for $m = 4, 5, ..., 8$. Each model's bases are plotted both within the 3D sphered color space (Fig.~\ref{fig:sc_models}a) and as a projection onto the chromatic plane (Fig.~\ref{fig:sc_models}b).

In the $m=4$ model, two basis vectors lie mostly within the chromatic plane and are aligned with red and yellow-green, roughly corresponding to two dominant heavy-tailed directions seen in the data distribution (Fig.~\ref{fig:stats}e). The remaining two vectors are elevated along the achromatic axis toward black and white, but retain a chromatic component in the blue direction, as indicated by their slight tilt from vertical. The tilt toward blue is needed as this region of hue space would otherwise remain unrepresented by the other two vectors within the chromatic plane.

When the number of basis vectors increases to five, a third basis vector is placed within the chromatic plane oriented toward the blue direction.
This placement is expected as the data exhibits a heavy-tailed extension in this direction, in addition to the red and yellow-green directions which are represented by two other basis vectors. 
The other two basis vectors now align closely with the achromatic axis, providing an explicit description of luminance variation. 
Thus, the five-basis-vector model factorizes chromatic and achromatic structure into distinct groups of basis vectors: three basis vectors tile the chromatic plane and two basis vectors encode positive and negative variations along the achromatic axis. 

\begin{figure}[!htb]
    \makebox[\textwidth][c]{\includegraphics[width=8.68cm]{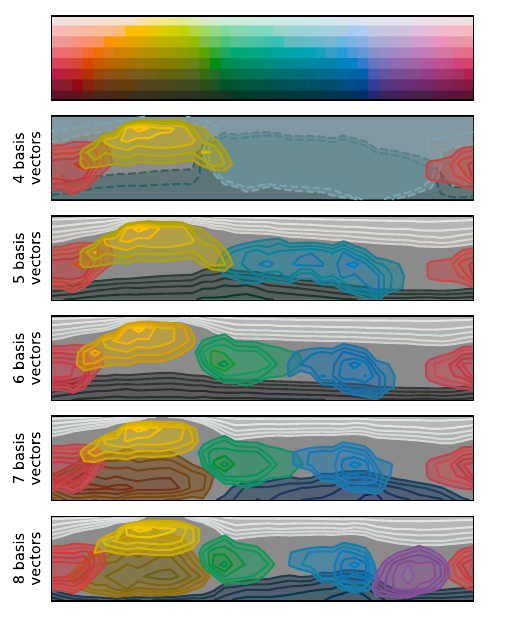}}
    \caption{
         Response maps of latent variable MAP activations ${s}_i$ as a function of 2D location within the Munsell color chart (top). Each panel shows the iso-response contours for each ${s}_i$ within a particular sparse coding model, with $m$ increasing top to bottom. Solid contour lines indicate response levels from 55\% to 95\% of each latent variable’s maximum, in 10\% increments.  Contour color indicates the direction of the corresponding basis vector $\mathbf{a}_i$ in sphered color space.
         The four-basis-vector model places strong weight on its two largely achromatic basis vectors to describe the blue/green directions. This is due to their opposing nature and the lack of other basis vectors that could be used to describe this region of color space. To show that these vectors are indeed sensitive to luminance variation, their iso-response contours are shown at 20\% and 30\% of each coefficient's maximum in dashed lines. 
    }
    \label{fig:munsell}
\end{figure}

At $m=6$, a fourth vector emerges within the chromatic plane, and the black and white vectors remain aligned to the achromatic axis. The new basis vector within the chromatic plane is aligned to green, nudging the previously yellow-green vector toward yellow, and the bluish vector toward the center of blue.  This nudging of the other vectors arises from the dynamics of adaptation in the sparse coding model, whereby bases both cooperate and compete to form an optimal tiling of the data distribution.  The resulting arrangement---vectors aligned to red, green, blue, and yellow---forms a strong correspondence with the unique hues. Psychophysical characterizations of the unique hues indicate that they do not typically align with the cone-opponent axes, but instead (with the exception of red) often lie in the quadrants between them~\cite{webster2000variations,wuerger2005cone,malkoc2005variations,wool2015salience}. Moreover, color-opponent pairs (green/red and blue/yellow) are not necessarily collinear (e.g., the line between red and green does not pass through the origin). 
The adapted $m=6$ model exhibits these properties: the hue-sensitive basis vectors do not necessarily align with the cardinal cone-opponent directions, and the opponent pairs are not collinear. 
Note that subjective unique hue settings vary considerably across observers and data sets~\cite{kuehni2004variability,malkoc2005variations} and hue angles depend on the scaling of the cone-opponent axes. For this reason, we focus on the broader geometric correspondence rather than a precise match to any single set of unique-hue measurements.
This general motif is maintained when adapting $m=6$ models to each individual dataset (Fig.~S3).  It should be emphasized that, given the wide variety of ways six basis vectors could tile 3D space, it would be highly unlikely for this particular solution to emerge by chance. This result is examined more extensively below.

At $m=7$, the basis vector aligned with black appears to split into two, each with a slight hue component, resulting in two vectors aligned to darker colors (i.e., brown and dark blue/purple). The other vectors remain largely unchanged from the $m=6$ solution, albeit with the yellow and blue vectors nudged slightly toward white.  
% an additional basis vector emerges near the yellow direction but with a negative elevation angle, suggesting sensitivity to similar hue content but with reduced achromatic (i.e., brown). Luminance-selective vectors remain present, with one shifting slightly off the vertical axis.

Finally, in the $m=8$ model, an additional vector emerges within the chromatic plane aligned with violet/purple, enabling the dark blue/purple vector from $m=7$ to revert to black. The directions of the other $m=7$ vectors remain mostly unchanged. 

Taken together, these models illustrate a progression in how the sparse coding basis adapts to the structure of the LMS distribution. All of the models form complete representations of the three-dimensional space, and all are adapted to the statistics of the data. 
They differ, however, in how parsimoniously they tile the heavy-tailed structure of the data. 
The four-vector model provides a minimal but awkward fit, requiring the achromatic vectors to participate in representing chromatic variation. 
The five-vector model gives a more natural separation between chromatic and achromatic structure, with three basis vectors spanning the chromatic plane and two aligned with the achromatic axis. 
The six-vector model is distinguished by the emergence of four chromatic vectors aligned with red, green, blue, and yellow, together with two achromatic vectors aligned with black and white. 
As shown below, this arrangement is unique among the other learned bases in producing three mutually exclusive opponent pairs. Models with seven or eight vectors further subdivide this representation, suggesting finer-scale chromatic or luminance-related subcategories rather than a qualitatively new opponent organization.

% Two key features emerge from these results. First, the bases tend to bifurcate between luminance-aligned (elevation $\pm90$ deg.) vs. chromatically-aligned (elevation $0$ deg.) directions (with the exception of the brown vector appearing at $m=7$ and $m=8$). This division is not enforced by constraints but instead arises from the statistics of the data. Second, the hue-selective basis vectors populate the chromatic plane in a structured sequence as more vectors are added, reflecting an efficient tiling of the chromatic plane. 

While Figure~\ref{fig:sc_models} reveals how the basis vectors $\mathbf{a}_i$ are arranged in the sphered color space, the manner in which color is encoded by these vectors requires examining how the corresponding latent variables $s_i$ activate in response to stimuli placed at different points in the space.  
These activations are computed via the MAP-inference of equations~\ref{eqn:energy} and \ref{eqn:nonneg_constraint}, and hence are
nonlinear functions of the stimulus vector $\mathbf{x}$.
% , and exhibit both cooperative and competitive interactions (as explained later).  
% it remains difficult to directly and intuitively assess their chromatic tuning. 
Visualizing these response functions, ${s}_i(\mathbf{x})$, within the 3D sphered color space is problematic, so we instead plot each latent variable's response as a function of its 2D location within the
Munsell color chart, a color standard used in cross-cultural linguistic studies of color naming. These plots are shown in Figure~\ref{fig:munsell}.
% , reveal the color tuning of the latent variables, $s_i$, that emerges due to the arrangement of the corresponding basis vectors, $\mathbf{a}_i$.
% [moving this sentence to beginning of next paragraph] The progression of tuning from broad to narrow as basis vectors are added to the model mirrors the increasingly refined tiling of the chromatic plane seen in Figure~\ref{fig:sc_models}. 

The progression of color tuning from broad to narrow as basis vectors are added to the model mirrors the increasingly refined tiling of the chromatic plane seen in Figure~\ref{fig:sc_models}. 
% The sequence in which the sparse coding models tile the Munsell color chart 
It is also reminiscent of the manner in which color terms are added to a color vocabulary. When additional vectors are added to the model, they 
% either encode a new region of the chart or 
subdivide a region that was previously represented by a single vector. For instance, the transition from five to six basis vectors introduces a new basis vector that separates blue and green, analogous to how languages with five basic color terms might introduce a distinct term for green or blue to resolve prior ambiguity.
Moreover, earlier-acquired basis vectors tend to persist in models with more vectors. For example, vectors that represent white, black, red, blue, yellow, and green are largely preserved in the seven- and eight-vector models.

\subsection*{A sparse coding account of the unique hues}

\begin{figure}[!t]
    \makebox[\textwidth][c]{\includegraphics[width=8.7cm]{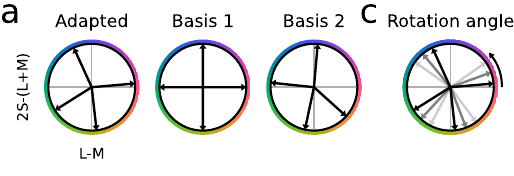}}
    \makebox[\textwidth][c]{\includegraphics[width=8.7cm]{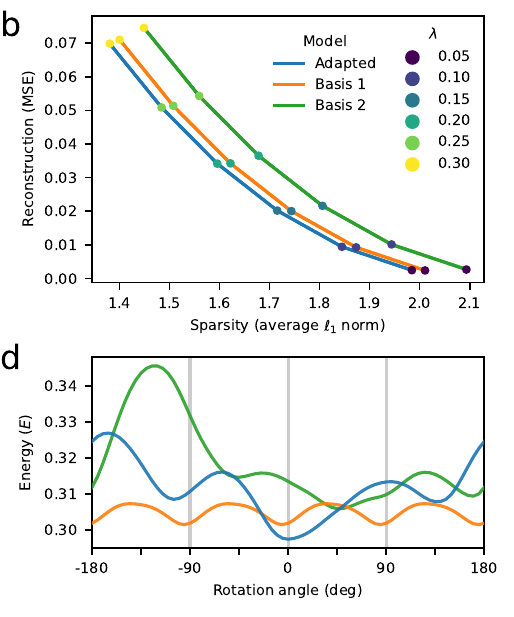}}
    \caption{
        Comparison of the adapted 6-basis-vector model with alternative models containing six basis vectors. 
        \textbf{a}) 
            The three bases analyzed, projected onto the chromatic plane. All bases span the full 3D space and include two achromatic vectors (white and black). Left to right: the adapted basis, a cardinal cone-opponent basis, and a teal-lime-orange-purple basis.
        \textbf{b})     
            Reconstruction quality (mean squared error) versus sparsity (average $\ell_1$ norm) curves for each basis, sweeping across sparsity thresholds $\lambda$. Across thresholds, the adapted basis consistently achieves lower MSE and greater sparsity.
        \textbf{c}) 
            Each basis is rotated in azimuth (about the vertical axis), altering the hue tuning of the chromatic vectors. 
        \textbf{d}) 
            Energy (the sum of MSE and $\ell_1$ sparsity weighted by $\lambda$) as a function of azimuth rotation (see \textbf{c}). The minimal-energy configuration occurs for the adapted basis at zero rotation.
        }
        \label{fig:why_adapted}
\end{figure}

Among the models discussed in the previous section, the 6-basis-vector model holds particular interest due to its alignment with the unique hues. Given the extensive literature surrounding these hues and their foundational status in color science, we study this model in detail. In doing so, we demonstrate why this basis is useful for representing the statistics of color in natural scenes, and we draw similarities between the properties of this model and the phenomenology of the unique hues.

To understand why the 6-basis-vector model arrives at its solution, it is important to consider how the sparse coding model represents data. 
In the model, any data point $\mathbf{x}$ is represented by scaling each unit-norm vector $\mathbf{a}_i$ according to its latent variable $s_i$ and then adding them together.  Graphically, this is equivalent to positioning the rescaled vectors end-to-end to reach the data point. Thus, the sum of latent variables (sparsity penalty in Eq.~\ref{eqn:energy}) can be interpreted as the total path length traversed by basis vectors to describe a data point. 
The puzzle solved by the sparse coding model is how to arrange the basis vectors so that, when averaged over the distribution of the data, the total path length is minimized. 
In doing so, the model finds a configuration that efficiently spans the LMS activation distribution.  The emergence of vectors resembling the unique hues can thus be understood as a consequence of the structure of the data and these model assumptions.

To demonstrate and verify this intuition quantitatively, we compare the efficiency and reconstruction quality of the unique-hue basis to those of two alternative 6-vector bases (Fig.~\ref{fig:why_adapted}a). One alternative is a canonical cone-opponent basis aligned with the axes of the sphered color space (basis 1). The other is a basis roughly aligned with teal, lime, orange, and purple (basis 2), following~\cite{bosten2014empirical} which used these hues to demonstrate that human subjects can use such a basis to describe color, and because they have served as a subject of debate with respect to the unique hues~\cite{saunders1997there,broackes1997could}. Note that both alternative bases retain two achromatic vectors (black and white), ensuring a fair comparison in the full 3D space.

Each model's performance is evaluated by sweeping over a range of values for sparsity parameter $\lambda$ and plotting the joint reconstruction error (mean-squared error; MSE) and average $\ell_1$ norm of latent variables. As shown in Figure~\ref{fig:why_adapted}b, the adapted basis achieves consistently lower MSE and lower $\ell_1$ norm across different settings of $\lambda$, indicating a more efficient representation under the model assumptions.

To further evaluate how this efficiency depends on the specific orientation of basis vectors within the chromatic plane, each basis is rotated in azimuth (Fig.~\ref{fig:why_adapted}c), and the total energy $E$ (Eq.~\ref{eqn:energy}) is computed as a function of rotation angle. As shown in Figure~\ref{fig:why_adapted}d, the adapted basis reaches a minimum at zero rotation, providing further confirmation that it is better matched to the structure of the data in comparison to a wide variety of alternatives.

% Note that $L_1$ sparsity in Figure~\ref{fig:why_adapted}b is evaluated because of the factorial exponential prior assumed over the nonnegative coefficients. The form of this prior is an assumption, and other well-motivated priors have the potential to change the results. This matter and potential future directions is discussed in more detail in the discussion. 

\subsection*{Emergence of color opponency} \label{sec:inference}
\begin{figure}[t]
    \centering
    \makebox[\textwidth][c]{\includegraphics[width=13.36cm]{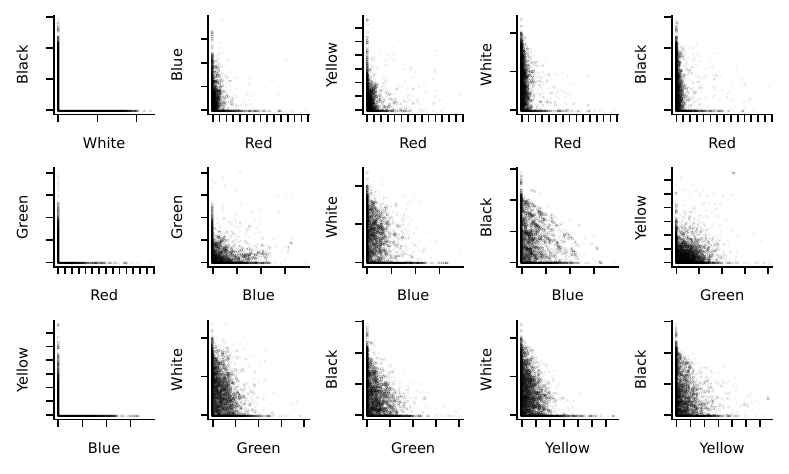}}
    \caption{
        Mutual exclusivity between opponent latent variables. Scatter plots show pairwise activations of inferred coefficients from the 6-basis-vector sparse coding model for 10,000 randomly sampled LMS activations. Each axis corresponds to the activation of a latent variable associated with the indicated color direction. Opponent pairs (red–green, blue–yellow, and black–white) exhibit mutual exclusivity due to sparse inference, whereas non-opponent pairs can coactivate. 
    }
    \label{fig:joint_scatter}
\end{figure}

\begin{figure*}[t]
    \centering
    \makebox[\textwidth][c]{\includegraphics[width=16.03cm]{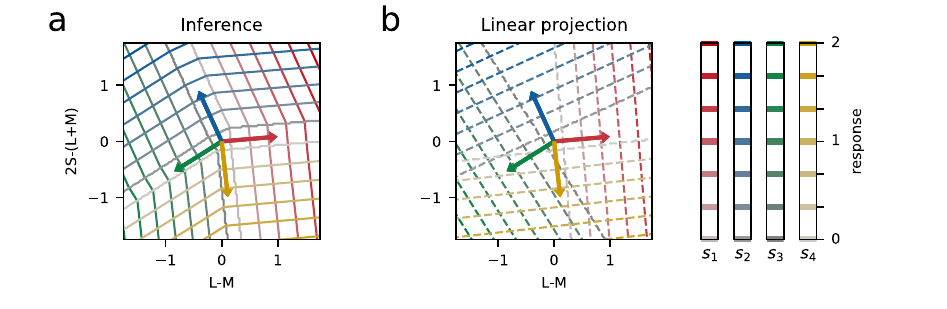}}
    \makebox[\textwidth][c]{\includegraphics[width=16.03cm]{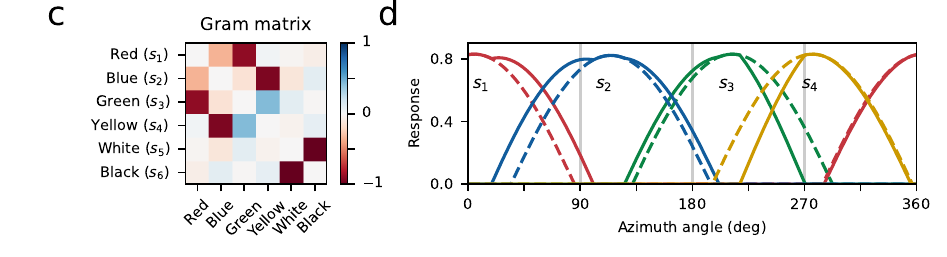}}
    \caption{
            \textbf{a}-\textbf{b}) 
                Iso-response contours (lines) colored by the corresponding basis vector. The hue saturation of the lines is set according to response magnitude (color bars at right). Contours are shown for both sparse inference (\textbf{a}) and linear projection  (\textbf{b}). 
                The linear projection representation is given by $s_i = g(\mathbf{a}_i^T\mathbf{x})$, where $\mathbf{x}$ is a point in the sphered color space, and $g(\cdot)$ is an element-wise rectifying nonlinearity (as in Eq.~\ref{eqn:lca_nonlinearity}). 
                Representations formed by sparse inference and linear projection are numbered $s_1$-$s_4$ according to the azimuth angle of their basis vectors, from smallest to largest. 
            \textbf{c}) 
                Gram matrix showing inner products between basis vectors, with positive (inhibition) and negative (excitation) values.
            \textbf{d}) 
                Tuning curves of the inferred representation (solid lines) in response to stimuli varying only in hue angle, compared to a linear projection with rectification (dashed lines). 
                % Nonlinear inference introduces deviations from simple feedforward projections. 
    }
    \label{fig:lca_v_rl}
\end{figure*}

Opponent pairs of unique hues are regarded as being perceptually opposite (e.g., there is no color that is simultaneously both red and green). 
% For example, the well-known CIELAB color space defines the axes of the isoluminant chromatic plane ($a^{\ast}$ and $b^{\ast}$) to be green/red and blue/yellow, respectively.
However, opponent pairs of unique hues are not necessarily opposite to one another in the cone-opponent plane (i.e., collinear) \cite{de1997hue,webster2000variations,wuerger2005cone,wool2015salience}. This raises the question of why these directions in color space are perceived as opposites, and what mathematical principles could reproduce such an opponent organization.

We find that inference in the sparse coding model gives rise to mutual exclusivity between latent variables which represent perceptually opponent colors, as well as the achromatic (white and black) latent variables (Fig.~\ref{fig:joint_scatter}). 
Figure~\ref{fig:lca_v_rl}a illustrates how inference represents stimuli within the chromatic plane by plotting iso-response contours of the chromatic latent variable representation.
The nonlinear interactions between latent variables $s_i$ in the LCA inference dynamics (Eq.~\ref{eqn:lca_dynamics}) cause their iso-response contours to curve, producing a complete and evenly distributed encoding of the chromatic plane without any overlap between iso-response contours belonging to an opponent pair (Fig.~\ref{fig:lca_v_rl}a). 
% of the competitive dynamics during inference (Eqs.~\ref{eqn:lca_dynamics} and \ref{eqn:lca_nonlinearity}). 

The interactions between latent variables during inference are mediated by the Gram matrix of inner-products between basis vectors $\mathbf{G}_{ij}=\mathbf{a}_i^T\mathbf{a}_j$ (Eq.~\ref{eqn:lca_dynamics}). 
Positive entries in the Gram matrix lead to inhibitory interactions: when two basis vectors are similar, activity in one's latent variable suppresses the other. For example, the green and yellow basis vectors have strong positive overlap in the Gram matrix (Fig.~\ref{fig:lca_v_rl}c), leading to curved iso-response contours that bend away from each other (Fig.~\ref{fig:lca_v_rl}a). 
This inhibitory interaction also ensures that pairs of basis vectors which represent opponent colors are mutually exclusive, as the contours of opponent pairs never cross despite a lack of collinearity or orthogonality.

Interestingly, negative entries of the Gram matrix (e.g., between blue and red) result in excitatory interactions between pairs of basis vectors. This interaction causes basis vectors to be recruited to represent regions of the chromatic plane despite the points within those regions having a negative projection on the basis vector. This prevents regions of input space from being represented by a single basis vector and thus enables a more even and complete tiling of the chromatic plane.

Somewhat paradoxically, opponent-pairs of basis vectors have large negative inner products, thus leading to an excitatory interaction. 
% It seems paradoxical that these pairs would have excitatory interactions yet still exhibit mutual exclusivity. 
How is it that they remain mutually exclusive?
The answer lies in the sparsity penalty. Inference selects the sparsest combination of active units to explain the data. Since simultaneous activation of opponent units would decrease sparsity, only one member of an opponent pair is used at a time.

However, the excitatory interactions between opponent pairs point to a broader issue: the model assumes a factorial (independent) prior over coefficients, but the empirical mutually exclusive distribution of inferred coefficients is not factorial (see Fig.~\ref{fig:joint_scatter}). This emerges from the interaction between sparsity and the structure of the data. A direction for future work is to explicitly incorporate this mutual exclusivity into a non-factorial prior (see Discussion).

Importantly, mutual exclusivity among opponent units is not imposed by the model but instead emerges naturally from inference in the generative model (Eq.~\ref{eqn:generative_model}).
Not all nonlinear transformations exhibit this property. Consider, for example, a representation generated by linear projection followed by rectification. Because responses in such a model are computed purely through linear projection, the iso-response contours in Figure~\ref{fig:lca_v_rl}b form straight lines perpendicular to their corresponding basis vectors.
% To produce a color-opponent representation given the $m=6$ basis, the weights on the basis vectors must interact. 
% Some form of interaction between the basis vector weights is necessary to produce a color opponency. 
% In Fig.~\ref{fig:lca_v_rl}b we consider a rectified linear projection and show that it is insufficient for producing color opponency. 
% it would be problematic to construct a representation which exhibits mutual exclusivity between elements of an opponent pair via simple linear projection onto the basis vectors. 
% Because the response is computed via linear projection, the 
% iso-response contours in Fig.~\ref{fig:lca_v_rl}b form straight lines perpendicular to the direction of their corresponding basis vector. 
The lack of orthogonality among the basis vectors leads to regions of the chromatic plane that are represented by combinations of hues that are considered perceptually opposed. 
For example, red and green units both respond in parts of the fourth quadrant, and blue and yellow responses overlap in the third quadrant. There are also significant regions of the chromatic plane encoded by only the blue vector or only the red vector, leading to metamerism: many distinct chromatic stimuli yielding identical responses. These problems cannot be solved by simply increasing the threshold (which would exacerbate metamerism), or introducing alternative pointwise nonlinearities.

Finally, because inference forms three mutually exclusive pairs of latent variables, they define a perceptually parsimonious $\mathbb{R}^3$ space in which each axis corresponds to a red-green, blue-yellow, or black-white opposition (Fig.~\ref{fig:overview}c). This structure reflects the phenomenological organization of color experience and is unique to the 6-basis-vector model. The emergence of these three opponent pairs is not a generic consequence of using six basis vectors, as this property depends on the specific orientations that the model learns from the data. For instance, a model with six basis vectors could instead represent three chromatic directions, one achromatic-increment direction (e.g., white), and two distinct decrement directions (e.g., black and brown); in such a case, pairwise mutual exclusivity would not occur, as white would be opponent to both black and brown. The other models also fail to construct a three-pair mutually exclusive opponent-color space: the four-vector model contains no opponent pairs, the five-vector model only one (black-white), and the seven-vector model introduces deviations from pairwise opponency, such as white being mutually exclusive with both black and brown.
Thus, the 6-basis-vector model uniquely combines numerical sufficiency with statistically driven directions to recover a perceptually meaningful, 3D opponent-color space.

\subsection*{Model tuning curves resemble perceptual hue scaling functions}

A well-established method for probing color appearance is hue scaling, in which participants rate the perceived proportions of a set of colors for a battery of colored stimuli that tile a color space \cite{de1997hue,bosten2014empirical}. To draw parallels with such psychophysical characterizations of the unique hues, we examined the responses of the 6-basis-vector model to stimuli that vary only in azimuth angle within the chromatic plane of the sphered color space.

Figure~\ref{fig:lca_v_rl}d shows the tuning curves of the inferred sparse coding representation to unit-norm chromatic stimuli sweeping through azimuth. For comparison, the tuning curves of the same basis under a linear projection (i.e., $g(\mathbf{A}^T \mathbf{x})$ as seen in Fig.~\ref{fig:lca_v_rl}a) are also plotted (dashed lines).
Under linear projection, all curves have the same shape and differ only by phase shifts. By contrast, the inferred representation (solid lines) exhibits non-uniform tuning widths due to interactions between units during inference. For example, excitatory interactions broaden responses near $60^\circ$, while inhibitory suppression narrows responses around $240^\circ$. These interactions produce red and blue tuning curves that are broader than those of green and yellow. 
This resembles the asymmetry reported by De Valois \textit{et al.}~\cite{de1997hue}, in which blue and red were used to describe a broader range of azimuth angles than green and yellow in human hue-scaling judgments. Thus, the correspondence is not only in the learned basis directions, but also in how sparse coding inference combines those directions to describe intermediate hues.

\section*{Discussion}\label{sec:discussion}

% The sparse coding framework introduced here not only models the distribution of naturalistic LMS activations, but also offers insight into how biological visual systems might represent color in a manner that aligns with perception.

The two main contributions of this paper are (1) the finding of strongly non-Gaussian structure of LMS activations in response to a large database of natural images, and (2) showing that a linear generative model adapted to this data aligns with the unique hues. 
A key observation enabling the second contribution is the asymmetry in the data, particularly within the chromatic plane, which motivates the use of a nonnegative code for its representation.
Nonnegativity, when combined with sparse coding inference, introduces mutual exclusivity between coefficients and yields representations that mirror human perceptual judgments.
It is remarkable that not only do the unique hues themselves emerge from the model, but the structure of their interactions also arises from fitting a single, linear generative model, with few other assumptions, to the data.

\subsection*{Implications and model predictions}

When framed as a probabilistic generative model~\cite{olshausen1997sparse}, the sparse coding model can be seen as constituting a prior over color appearance acquired from exposure to the natural environment.  
This raises the question of whether one can find evidence of the influence of this prior in judgments of color appearance, similar to how priors have been shown to influence judgments in other tasks~\cite{griffiths2006optimal}.  
For example, given a noisy or uncertain observation of a color, the model would predict an observer's estimate of the color to be biased toward regions of higher probability in the prior.  
% However, as noted previously, it may be worth modifying the prior over the latent variables to include coupling terms among latent variables in order to make a better approximation to the true distribution.  
Alternatively, since the sparse coding model is an approximation, and the full color distribution is relatively low-dimensional, it may be feasible to use the empirical distribution (Fig.~\ref{fig:stats}) directly to test whether it predicts color judgments under uncertainty. 

% We can view the energy function in equation~\ref{eqn:energy} as the negative log of the joint distribution over data and latent variables, P(\mathbf{x},\mathbf{s}).  Therefore integrating over \mathbf{s} give the probability of the model generating a given color, x, via p(x) = \int_s p(x|s)\,p(s) ds.  In this sense, p(x) serves as a prior over the distribution of color in natural scenes.  

Another question that arises from the sparse coding model is whether one should expect individual neurons in the visual system to exhibit tuning to the unique hues similar to that of the latent variables of the model (Fig.~\ref{fig:munsell}).  
This question is difficult to answer at present because we lack a theory for how space and color should be jointly represented. 
A widely held view is that the visual system aims to disentangle image data into independent factors of variation such as illumination, object shape, and reflectance properties~\cite{adelson1996perception}. Previous work applying sparse coding or ICA to color natural image patches recovers spatial Gabor-like basis functions with color-opponent structure~\cite{wachtler2001chromatic,doi2003spatiochromatic,hyvarinen2009natural}. However, information about color, shape and illumination still remains entangled in such a representation since the latent variables co-vary with each of these properties.  
We conjecture that disentangling color and shape will require more complex, multistage models that have the ability to factorize components~\cite{frady2020resonator} as opposed to the simple type of linear generative models explored here and elsewhere. 
% It may be that the latent variables of these models are effectively represented and manipulated internally not by single neurons but via high-dimensional neural populations, as in Vector Symbolic Architectures~\cite{kanerva2009hyperdimensional,kleyko2022vector}, which could explain why the search for neurons tuned to the unique hues has proven a fraught endeavor~\cite{conway2023color}.
It may be that the latent variables of these models are effectively represented and manipulated internally not by single neurons but via high-dimensional neural populations, as in Vector Symbolic Architectures~\cite{kanerva2009hyperdimensional,kleyko2022vector}, which could explain the failure to find neurons tuned to the unique hues~\cite{conway2023color}.

\subsection*{Relation to prior work}

Other multistage models of chromatic information processing have also been proposed. 
Doi \textit{et al.}~\cite{doi2003spatiochromatic} applied a similar three-stage model to spatiochromatic natural images, which were sampled by a realistic cone mosaic. The model consisted of a sphering step followed by ICA, which were both adapted to the structure in the spatiochromatic natural image patches. The ICA model derived cone-selective receptive fields; however, it is unclear to what extent these filters relate to color perception. 
De Valois and De Valois~\cite{de1993multi} demonstrated how the visual system could transform color representations in a multistage model proceeding from LMS to cone-opponent to red-green/blue-yellow color opponent dimensions, given biophysical constraints such as relative cone ratios and spatially symmetric center-surround receptive field structure in the retina.
Here we show how the specific computations in these stages can be derived within a first-principles framework applied to image statistics.

Yendrikhovskij~\cite{yendrikhovskij2001computing} applied clustering algorithms to pixels from natural images in a perceptually uniform color space and found that the resulting clusters aligned with so-called `universally named' colors.
However, later studies struggled to replicate this result~\cite{steels2005coordinating,zaslavsky2020communicative}. Given the structure of the data shown in Figure~\ref{fig:stats}c-e, it is not clear that a clustering model per se is appropriate for capturing the structure of natural color statistics (see also below).
% \cite{zaslavsky2020communicative}. 

Philipona and O’Regan~\cite{philipona2006color} provide an alternative account of the unique hues based on the observation that certain surface reflectance spectra, when filtered through the cone absorption functions, remain relatively invariant under changes in illumination. They show that these singularities in color space, which would be most informative of surface reflectance, correspond to the unique hues. This finding suggests that the asymmetric and heavy-tailed structure in the LMS distribution reported here may be related to the relative invariance of certain reflectances to illumination. Investigating this relationship would be an interesting direction for future work.

Webster and Mollon~\cite{webster1997adaptation} found considerable variation in natural-image chromaticities along a blue-yellow direction, with greater variation along S-(L+M) than along L-M. This broad anisotropy has been replicated here and elsewhere~\cite{ruderman1998statistics}. Specifically, they found that the dominant axis of chromatic variation was not aligned with the cone-opponent axes. Direct comparison between their color distribution and the one reported here is limited by two differences. First, each image in our dataset was rescaled by its mean cone response (see Methods), whereas their analysis did not include an analogous normalization. Second, their post-receptoral space differs slightly from the PCA-derived post-receptoral space used here and, more importantly, uses different scalings of the chromatic axes. They rescale the axes according to estimated detection thresholds, whereas we rescale the axes to have unit variance. Nonetheless, the sparse-coding models learn basis vectors aligned with the blue-yellow directions (Fig.~\ref{fig:sc_models}). This off-axis bias could reflect scene content, as both datasets contain mixtures of foliage and sky. Understanding how the non-Gaussian structure of the LMS distribution relates to image content remains an important direction for future work.

\subsection*{Bases vs. categories}

The unique hues are in some cases regarded as discrete color categories (as in color naming) and in other cases as a basis for representing the full continuum of colors.  The sparse coding model is an example of the latter.
Whereas categorization produces a discrete code that assigns many points in stimulus space to a single category, the sparse coding model produces a continuous code by linearly combining basis vectors to describe any point in color space, potentially without loss.  
However in the extreme case where sparsity is so heavily encouraged that only a single basis vector is used to describe a given point in color space, it becomes equivalent to a vector quantization model similar to categorization or clustering (see~\cite{hinton1997generative} for a useful discussion of the relation between sparse coding and clustering).  Previous attempts to measure categorical effects in color perception have produced mixed evidence for categorical boundaries consistent with the unique hues~\cite{witzel2018red}. It is not clear that the sparse coding model would produce such categorical effects, especially if the prior were to be modified to account for facilitatory interactions between latent variables (see below).  

Interestingly, the sparse coding model also captures the atomistic property of the unique hues originally noted by Hering, since a color aligned with any one basis vector would be described by that vector alone, whereas a color lying between basis vectors would be described as a linear combination of them.

% \cite{witzel2018red}, and in other cases treated as perceptual reference points in a continous color space \cite{de1997hue,webster2000variations}.
% ; the relationship to color naming is therefore unsettled \cite{saunders1997there}. 
% The theory we propose here is in line with the notion of the latter. Moreover, it mathematically formalizes how the unique hues serve as a basis for describing all colors via linear combination in a model derived from the statistics of the natural environment. 

% Other theoretical approaches to understanding color naming have leveraged an information bottleneck framework for efficiently encoding color space into a lexicon~\cite{zaslavsky2018efficient}. Extensions of this approach have also incorporated image statistics \cite{zaslavsky2020communicative}. 

\subsection*{Model extensions and future work}

There are natural extensions to the sparse coding model.
First, while we assume a factorial prior over the latent variables, the inferred representations clearly deviate from this assumption. This is most notable through mutual exclusivity between opponent-color units. Future work could explore incorporating structured priors that explicitly model mutual exclusivity. 
Second, in this work we compute the latent variables via MAP estimates rather than
sampling from the full posterior distribution. 
A disadvantage of this approach is that these estimates are biased toward zero due to the sparse prior, requiring the basis function norms to be constrained to one (otherwise they would grow without bound). By instead sampling from the posterior (e.g., via Langevin sampling), the model could potentially learn the basis vector norms along with parameters of the prior, such as excitatory and inhibitory couplings between latent variables~\cite{fang2022learning}.

Lastly, cone spectral sensitivities vary greatly across the animal kingdom~\cite{osorio2008review}. Prior work has asked whether primate cone sensitivities are optimized for sampling natural spectra~\cite{lewis2006cone}. The results presented here are all based on spectral sensitivities of the trichromatic primate visual system; however, this theoretical approach is general and can be applied to any species with spectrally distinct photoreceptors. 
In this sense, cone sensitivities define the photoreceptor basis through which spectra are sampled, whereas the sparse-coding model addresses how the resulting cone responses may be organized into a post-receptoral code. Human spectral sensitivities vary across observers, yet unique-yellow settings can remain remarkably stable~\cite{neitz2002color}. This raises the possibility that the higher-order structure captured here is robust to some variation in cone spectral sensitivities, a question that could be tested by repeating the present analysis under different photoreceptor sensitivities. More broadly, applying the same framework across species could test whether similar coding principles operate in visual systems with very different photoreceptor bases.
Recent findings 
% , independent of our own, 
on color vision in \textit{Drosophila} have revealed a very similar coding strategy to what we have presented here: photoreceptor activations are initially decorrelated via horizontal cells~\cite{heath2020circuit} and then a later stage is believed to use a recurrent nonnegative encoding strategy, similar to LCA, to produce hue-selective neurons with curved iso-response contours similar to those in Figure~\ref{fig:lca_v_rl}b~\cite{christenson2024hue}. 
To what extent these hue tuning properties can be explained by natural image statistics remains an open question. Nonetheless, these empirical findings bear a striking resemblance to the model proposed here, and their convergence offers the opportunity to test whether principles of efficient coding and perceptual inference can provide an account for color perception across the animal kingdom.

\section*{Methods}

\subsection*{Dataset of simulated LMS activations to natural images}

    \begin{table}[t]
        \centering
        \begin{tabular*}{\textwidth}{@{\extracolsep{\fill}}lccc}
            \toprule
            Dataset & Type & N. images & Location(s) \\
            \midrule
            Kyoto natural images \cite{doi2003spatiochromatic} & Calibrated LMS & 62  & Japan \\
            Barcelona calibrated images \cite{parraga2009new} & Calibrated LMS & 237 & Spain \\
            Ruderman, Cronin, and Chiao \cite{ruderman1998statistics} & Hyperspectral  & 12  & United States and Australia \\
            2022 NTIRE spectral recovery   \cite{arad2022ntire} & Hyperspectral  & 165 & Israel \\
            Foster and Reeves \cite{50hyperspectraldataset,foster2022colour} & Hyperspectral  & 27  & Portugal \\
            \bottomrule
        \end{tabular*}
        \captionsetup{width=\textwidth}
        \caption{
            Details of the natural image datasets used for simulating LMS cone activations. Each dataset is listed with its data type, number of images, and collection location(s).
        }
        \label{table:dataset}
    \end{table}    
    
    We constructed a large dataset of simulated LMS activations in response to natural images by combining five calibrated LMS and hyperspectral datasets, the details of which are provided in Table~\ref{table:dataset}. The collated dataset contains over 225 million LMS activations arising from 503 images. The motivation behind combining datasets is that it allows for a more diverse sampling of visual environments and reduces potential biases from any single camera, environmental setting, or selection bias of the investigator. The hyperspectral datasets were converted to cone responses using the ten-degree cone fundamentals from \cite{stockman2000spectral}. 
    
    We eliminated from the dataset any images that contain man-made objects.  We removed these images because the colors of man-made objects could inherit the very biases in human color perception that we seek to study, which would confound our results. That said, even scenes of only naturalistic objects could contain biases of what the photographer thought was worth photographing. It is hard to rule out this bias and could be present in any dataset. 
    Additionally, only images in the `naturalistic' category from the Barcelona dataset \cite{parraga2009new} were used, and the gray sphere in the left portion of each image was removed by cropping out the leftmost 200 pixels. We did not include the large simulated LMS cone natural image dataset of~\cite{tkavcik2011natural} as its PCA-summarized statistics were substantially different from the other datasets. Understanding the reasons for this and its implications deserves additional study.

    With the exception of the Kyoto dataset, all of the calibrated LMS datasets are proportional to linear cone absorbances and can be thought of as cone photon isomerizations, which will be nonlinearly transformed to cone outputs. Following Doi \textit{et al.}~\cite{doi2003spatiochromatic} (see also~\cite{wachtler2007cone}), the transform from absorbance $v$ to nonlinear response $r$ is modeled using an exponential saturation function $r=1-e^{-kv}$ as proposed by Baylor \textit{et al.}~\cite{baylor1987spectral}. Interestingly, this function describes the cumulative density function of the exponential distribution $p(v)=ke^{-k\,v}$, which is the maximum entropy distribution for a nonnegative variable $v$ with mean $1/k$. 
    We observed that cone absorbance distributions for a given image often resemble an exponential distribution, with $k$ varying with cone type and illumination level. Therefore, by setting $k$ equal to the inverse mean, $1/\langle v_i \rangle$, where $i \in \{L,M,S\}$ denotes a particular cone type, an approximately uniform distribution over $[0,1]$ will be produced on the cone outputs. (A similar line of reasoning was used by Laughlin to derive the optimal nonlinear response function for neurons in the fly visual system~\cite{laughlin1981simple}). Moreover, this choice of $k$ effectively applies von Kries adaptation to the images, in which each cone type’s response is rescaled independently according to its mean~\cite{wyszecki2000color}. Mathematically, the transformation from linear cone inputs $v_i$ to nonlinear outputs $r_i$ is as follows:
    \begin{align}
        r_i &= 1-\text{exp}\left(-\frac{v_i}{\langle v_{i}\rangle}\right).
    \end{align}
    Note that the Kyoto dataset uses a similar function to convert to nonlinear outputs; however, they set $k$ such that the median of $r_i$ is equal to 0.5 (see \cite{doi2003spatiochromatic} for details). As a final step of processing, the mean is subtracted from each image (computed over the spatial and channel dimensions), such that the mean over the entire dataset was zero prior to the redundancy reduction step, which is described in the subsequent section.

    \subsection*{PCA for redundancy reduction}
    
    Following Ruderman \textit{et al.}~\cite{ruderman1998statistics}, we apply PCA to remove second-order statistical redundancies from simulated LMS activations.
    
    Given an LMS cone activation vector $\mathbf{r} = \begin{pmatrix}r_L & r_M & r_S\end{pmatrix}^T$, we seek a linear transformation $\mathbf{W} \in \mathbb{R}^{3 \times 3}$ such that:
    \begin{align}
        \mathbf{x} &= \mathbf{Wr} \quad \text{with} \quad \langle \mathbf{xx}^T \rangle = \mathbf{I},
    \end{align}
    where $\langle \cdot \rangle$ denotes the expectation over the dataset, and $\mathbf{I}$ is the identity matrix.

    Such a linear transformation can be computed using PCA. The covariance matrix of the LMS inputs is decomposed as
    \begin{align}
        \langle \mathbf{r} \mathbf{r}^T\rangle = \mathbf{E}\bm{\Lambda}\mathbf{E}^T,
    \end{align}
    where $\mathbf{E}$ contains the eigenvectors (principal components) and $\bm{\Lambda} = \text{diag}(\sigma^2_1, \sigma^2_2, \sigma^2_3)$ is the diagonal matrix of their corresponding variances, sorted such that $\sigma^2_1 > \sigma^2_2 > \sigma^2_3$.

    To decorrelate and rescale the data to have unit variance in all directions, the data are first rotated to the principal component basis using $\mathbf{E}^T$, and then rescaled by the inverse standard deviation using $\bm{\Lambda}^{-1/2} = \text{diag}(\sigma^{-1}_1, \sigma^{-1}_2, \sigma^{-1}_3)$. Combined, this yields the linear transformation
    \begin{align}
    \mathbf{W} = \bm{\Lambda}^{-1/2}\mathbf{E}^T.
    \end{align}
    This maps LMS activations into the sphered color space.

    \subsection*{Locally Competitive Algorithm (LCA)}

    We solve equations~\ref{eqn:energy}-\ref{eqn:nonneg_constraint} using a nonnegative variant of LCA~\cite{rozell2008sparse}. LCA evolves subthreshold variables $u_i$ whose thresholded outputs $s_i$ are the sparse coefficients:
    \begin{align}
        \tau \dot{u}_i + u_i &= \mathbf{a}_i^T\mathbf{x} - \sum_{j\neq i}\mathbf{G}_{ij} s_j \label{eqn:lca_dynamics}\\
        s_i &= \max(0, u_i-\lambda) = g(u_i)\label{eqn:lca_nonlinearity}
    \end{align}
    where $ \mathbf{G}_{ij} = \mathbf{a}_i^T\mathbf{a}_j$. The rectifying nonlinearity $g(\cdot)$ enforces the $\ell_1$-sparsity penalty on the coefficients while imposing an infinite energy barrier at $s_i<0$, satisfying the nonnegativity constraint.  Letting the dynamics settle to a steady state yields the coefficients $\mathbf{s}$ that minimize $E$, corresponding to the MAP estimate under the model. The basis is learned by stochastic gradient descent on the energy $E$ with respect to $\mathbf{A}$, using the inferred coefficients  $\mathbf{s}$. This yields the following learning rule:
    \begin{align}
        \Delta \mathbf{a}_i = \eta \left[ \mathbf{x} - \sum_{j=1}^m \mathbf{a}_j s_j \right] s_i.
    \end{align}
    The update rate $\eta$ is set small so that the energy decreases continuously until convergence.
    After each update, the basis vectors are normalized to have unit norm: $\lVert\mathbf{a}_i\rVert_2 =1 \text{ for all } i$.

    The sparsity parameter $\lambda$ was adjusted during learning such that a constant signal-to-noise ratio (SNR) was maintained in the reconstruction error. The SNR is a function of mean squared error (MSE) where $\text{MSE} = \lVert\mathbf{x} - \mathbf{As}\rVert_2^2$. SNR in decibels (dB) is computed as follows,
    \begin{align}
        \text{SNR} &= 10 \log_{10} \text{MSE}^{-1}.
    \end{align}
    The sparsity parameter $\lambda$ was set such that 16 dB SNR was maintained for all the bases shown in this paper. We did not observe significant changes to the optimal bases when $\lambda$ enforced higher or lower SNRs. 

\backmatter
\vspace{-5pt}
\bmhead{Acknowledgements}
A.B. and B.A.O. are supported by Air Force Office of Scientific Research (AFOSR) MURI grant (FA9550-21-1-0230).
The authors gratefully acknowledge the feedback and support of the MURI team.
\vspace{-5pt}
\bmhead{Author contributions}
A.B., E.P.F., and B.A.O. designed research; A.B., B.A.O., and  E.P.F. performed research; A.B., E.P.F., and B.A.O. analyzed data; and A.B. and B.A.O. wrote the paper.
\vspace{-5pt}
\bmhead{Data and code availability}
% All data and code necessary to support the paper’s conclusions are present in the main text, supplemental information, or available from the corresponding author upon reasonable request.
The natural image databases underlying the results are available for download online. Code is available upon request. 
\vspace{-5pt}
\bmhead{Competing interests}
The authors declare no competing interests.
\vspace{-5pt}
\bmhead{Supplemental document}
See the supplemental document for supporting content and figures.

\bibliography{ref}

% \newpage
% \section*{Supplemental Information}
\begingroup
\setcounter{figure}{0}
\renewcommand{\thefigure}{S\arabic{figure}}
\renewcommand{\figurename}{Supplementary Fig.} % or "Supplementary Figure"
% (optional, if you also have tables in SI)
\setcounter{table}{0}
\renewcommand{\thetable}{S\arabic{table}}
\renewcommand{\tablename}{Supplementary Table}

\begin{figure}[t]
    \centering
    \makebox[\textwidth][c]{\includegraphics[width=16.98cm]{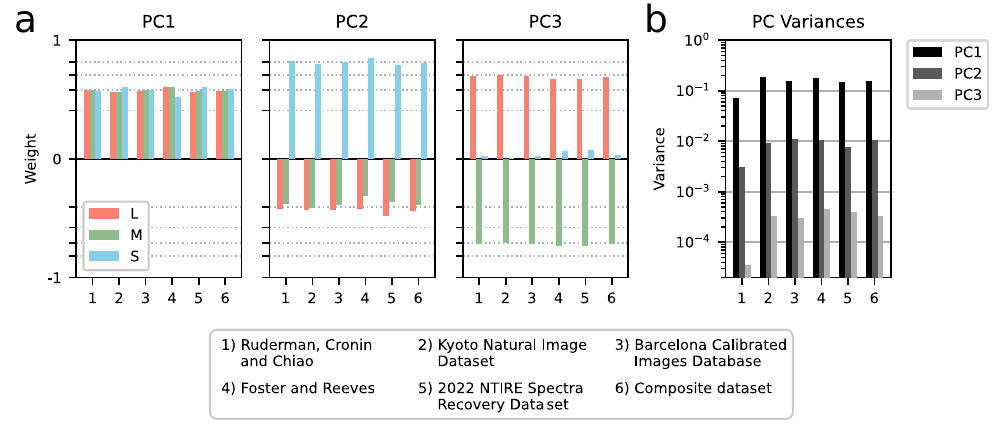}}
    \caption{
        \textbf{a}) Principal components and corresponding variances for each of the five constituent datasets. Horizontal dashed lines indicate
        $\pm\text{sqrt}(1/6)$, 
        $\pm\text{sqrt}(1/3)$, 
        $\pm\text{sqrt}(1/2)$,
        and $\pm\text{sqrt}(2/3)$, revealing the near-integer multiple relationship in the principal components weight L, M, and S cone inputs. 
        \textbf{b}) Variance on each principal component for across datasets. 
    }\label{fig:suppcs}
\end{figure}

\begin{table}[h]
    \centering
    \begin{tabular*}{0.75\textwidth}{@{\extracolsep{\fill}}lccc}
        \toprule
        & \multicolumn{3}{c}{Kurtosis (excess)} \\
        \cmidrule(lr){2-4}
        Dataset & L+M+S & 2S-(L+M) & L-M \\
        \midrule
        Kyoto natural images \cite{doi2003spatiochromatic} & -0.88 & 6.66 & 23.39 \\
        Barcelona calibrated images \cite{parraga2009new} & -0.60 & 4.19 & 27.04 \\
        Ruderman, Cronin, and Chiao \cite{ruderman1998statistics} & -0.36 & 1.49 & 4.31 \\
        2022 NTIRE spectral recovery \cite{arad2022ntire} & -0.52 & 2.80 & 25.21 \\
        Foster and Reeves \cite{50hyperspectraldataset,foster2022colour} & -0.66 & 3.83 & 20.44 \\
        Composite dataset & -0.60 & 4.30 & 27.03 \\
        \bottomrule
    \end{tabular*}
    \captionsetup{width=0.75\textwidth}
    \caption{
        Excess kurtosis was calculated after projecting each dataset onto the three axes of the sphered color space. Positive excess kurtosis indicates heavy-tailed (sparse) structure, whereas negative values indicate sub-Gaussian structure. All datasets exhibit sparse structure along chromatic axes and sub-Gaussian structure along the achromatic axis.
    }
    \label{table:dataset_kurt}
\end{table}

\begin{figure}[t]
    \centering
    \makebox[\textwidth][c]{\includegraphics[width=8.7cm]{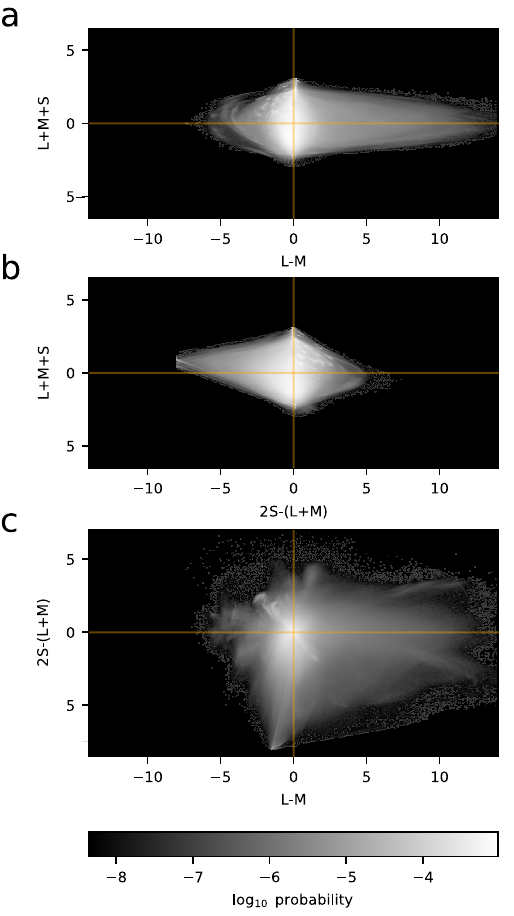}}
    \caption{
        Distribution of all 225,958,904 LMS activations from the dataset, plotted as a two-dimensional histogram after projection onto three different 2D planes defined by each possible pairing of axes in the sphered color space, as indicated by the axis labels in \textbf{a}-\textbf{c}.
    }
    \label{fig:full_density}
\end{figure}

\begin{figure}[t]
    \centering
    \makebox[\textwidth][c]{\includegraphics[width=17.8cm]{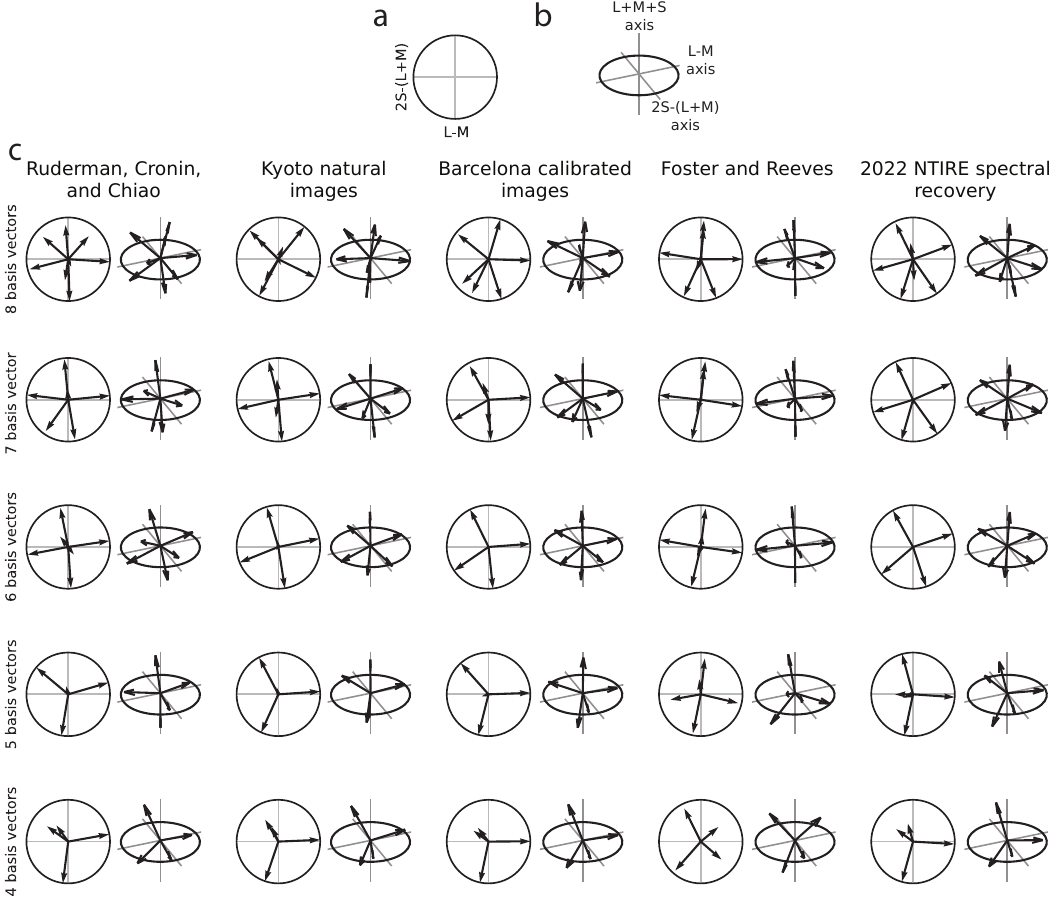}}
    \caption{
        Adapted nonnegative sparse coding bases from $m=4,5,...,8$ for each dataset (\textbf{c}). Each basis is shown in the 3D sphered color space and as a projection (axis labels shown in \textbf{b}) and onto the chromatic plane (axis labels shown in \textbf{a}).
   }
   \label{fig:sc_each_dataset}
\end{figure}

\endgroup

\end{document}